\documentclass[12pt]{article}
\usepackage{graphics}
\usepackage{psfig}
\usepackage{amssymb}

\newcommand{\rf}[1]{(\ref{#1})}
\newcommand{\beq}{\begin{equation}}
\newcommand{\eeq}{\end{equation}}
\newcommand{\bea}{\begin{eqnarray}}
\newcommand{\eea}{\end{eqnarray}}

\newcommand{\La}{\Lambda}

\renewcommand{\a}{\alpha}

\newcommand{\del}{\delta}

\newcommand{\oh}{\frac{1}{2}}

\newcommand{\ra}{\rangle}
\newcommand{\la}{\langle}
\newcommand{\prt}{\partial}
\newcommand{\mi}{\!-\!}
\newcommand{\equ}{\!=\!}
\newcommand{\pl}{\!+\!}

\newcommand{\cO}{{\cal O}}

\newcommand{\tW}{{\tilde{W}}}

\newcommand{\hZ}{{\hat{Z}}}

\newcommand{\SL}{\sqrt{\La}}

\begin{document}

\begin{center}
\vspace{24pt}
{ \large \bf The Geometry of ZZ-branes}

\vspace{30pt}

{\sl J. Ambj\o rn}$\,^{a,c}$
{\sl S. Arianos}$\,^{a,b}$,
{\sl J. A. Gesser}$\,^{a}$
and {\sl S. Kawamoto}$\,^{a}$

\vspace{24pt}
{\footnotesize

$^a$~The Niels Bohr Institute, Copenhagen University\\
Blegdamsvej 17, DK-2100 Copenhagen \O , Denmark.\\
{ email: ambjorn, arianos, gesser, kawamoto @nbi.dk}\\

\vspace{10pt}

$^b$~Dipartimento di Fisica Teorica, \\
Universit\`a di Torino, \& INFN-sezione di Torino\\
via P. Giuria 1, I-10125 Torino, Italy\\
{ email: arianos@to.infn.it}

\vspace{10pt}

$^c$~Institute for Theoretical Physics, Utrecht University, \\
Leuvenlaan 4, NL-3584 CE Utrecht, The Netherlands.\\

\vspace{10pt}

}
\vspace{48pt}

\end{center}

%\addtolength{\baselineskip}{0.20\baselineskip}
%\vspace{2cm}

\begin{center}
{\bf Abstract}
\end{center}

We show how non-compact (quantum 2d AdS) space-time emerges for  specific
ratios of the square of the 
boundary cosmological constant to the cosmological 
constant in  2d Euclidean quantum gravity.

\vspace{12pt}
\noindent

%\vfill

\newpage

\subsection*{Introduction}\label{intro}

In two beautiful papers Fateev, Zamolodchikov and 
Zamolodchikov (FZZ) and Zamolodchikov and Zamolodchikov (ZZ) 
quantized Liouville 
theory on the disk and on the pseudo-sphere \cite{fzz,zz}.
The boundary condition consistent with conformal invariance
on the disk is a generalized Neumann boundary condition which 
can be derived by adding a boundary cosmological constant 
to the Liouville action. From the explicit expressions
for correlation functions derived in \cite{fzz} one can 
reconstruct  many of results for non-critical string theories
and 2d quantum gravity derived earlier by matrix model techniques
\footnote{And of course the quantum Liouville theory 
contains a lot of additional information compared to the 
global, diffeomorphism invariant variables  accessible
by matrix model techniques.},
in particular the disk amplitude $W(Z,\La)$, which is 
a function of the boundary cosmological constant $Z$ and the
the cosmological constant $\La$. The pseudo-sphere is 
a non-compact space with no boundary and although one 
of course has to impose suitable boundary conditions ``at infinity''
in order to have a conformal field theory on the pseudo-sphere, there
is nothing like a boundary cosmological term. The word is 
not mentioned in \cite{zz}, {\it but the non-compactness of the 
pseudo-sphere is the foundation of the whole derivation}: one
uses the factorization 
\beq\label{0.0}
\la \cO_1 (x) \cO_2(y)\ra \to \la \cO_1(x)\ra \la \cO_2(y)\ra
\eeq
when the {\it geodesic distance} between $x$ and $y$ goes 
to infinity to derive an equation for certain operators.

In the modern string terminology boundary conditions of 
2d conformal field theories are almost synonymous to ``branes'',
and the work of FZZ and ZZ was instrumental to the recent 
development describing what is now called FZZ- and ZZ-branes
in non-critical string theory and their interpretation as 
eigenvalues in the matrix models and the relation to 
Sen's rolling tachyons. Martinec was probably 
the first to notice (or at least to state explicitly) 
the somewhat curious fact that the 
ZZ-branes could  be expressed as a ``difference'' between 
two FZZ-branes in the language of boundary states \cite{martinec}. This 
line of development culminated\footnote{While finishing the writing
of this paper a further elaboration of the work of 
SS has appeared \cite{ss.et.al}} with the paper of Seiberg and 
Shih (SS) \cite{ss} , where it was shown that (many of) the ZZ-branes were
directly related to the zeros of the disk amplitude $W_\La(Z)$.    
 
As emphasized above non-compactness of the pseudo-sphere is 
essential for the derivation of the ZZ-results.
However, what has been missing so far is any geometric understanding
of why these special ratios of $Z^2/\La$ are related to non-compact 
spaces. In fact, as we shall review, from the point of view of 
the disk amplitude, these values play (almost) no special role. 
{\it The purpose of this letter is to explain why the zeros of 
$W_\La(Z)$ indeed are related to non-compact geometries of the AdS type}.
We will for clarity confine ourselves to the simplest situation: 
``pure'' 2d quantum gravity, i.e.\ $c_{matter} \equ 0$ (or equivalently
$c_{Liouville}\equ 26$), 
where the geometric picture is most transparent. We will discuss 
the minimal $(p,q)$ model coupled to 2d quantum gravity in a separate 
paper \cite{aagk2}.

\subsection*{The disk amplitude}

Let us briefly discuss the disk amplitude of 2d quantum gravity
without coupling to matter fields, i.e.\ $c_{Liouville}
= 26$ in the corresponding  Liouville quantum 
theory. The framework of dynamical triangulations (or equivalently matrix 
models) provides a regularization of the path integral of the 2d quantum
gravity theory. The summation over geometries is replaced 
by the summation over triangulations constructed from equilateral 
triangles with side-length $a$. This lattice cut-off acts as 
the reparametrization invariant cut-off for the Liouville quantum
field theory. Denote the number of triangulations of the disk 
with  $n$ triangles and $l$ boundary links by $w_{n,l}$.
The generating function can be written as
\bea\label{1.1}
w(z,g)& =& \sum_{n,l} g^n z^{-l-1} w_{n,l}\\
&=&\oh \Big( z-gz^2+g(z-c(g))
\sqrt{(z-c_{+}(g))(z-c_{-}(g))}\Big) \nonumber
\eea
and was first found by Tutte in 1962 \cite{tutte}
by combinatorial arguments.
$w(z,g)$ is analytic in the complex $z$ plane except for
a cut $[c_{-}(g),c_{+}{g}]$. The functions $c,c_+,c_-$ are 
analytic functions of $g$ with a radius of convergence $g_c$.
The variables $z$ and $g$ can be related to the continuum 
boundary cosmological constant $Z$ and the continuum cosmological 
constant $\La$ by 
\beq\label{1.2}
g=g_c e^{-a^2 \La},~~~~z=z_c e^{aZ},~~~z_c=c_+(g_c),
\eeq
such that $g^n z^{-l}$ reproduces the Boltzmann weight of 
2d Euclidean gravity:
\beq\label{1.2a} 
{\rm Boltzmann~weight}= 
\exp\Big(-\La \int_D d^2\xi \sqrt{g} - Z \int_{\partial D} ds \sqrt[4]{g}\Big).
\eeq
Using the fact that 
\beq\label{1.3}
c(g)= z_c(1+\oh \a a \SL + \cdots),~~~c_+(g)=z_c(1-\a a \SL +\cdot),
\eeq
where $\a$ is a constant, 
one obtains (after a suitable rescaling of $Z$ and $\La$)
\footnote{We are using the same normalization as in \cite{ss}.}
\beq\label{1.4}
w(z,g) = \oh \Big(V'_{ns}+a^{3/2}W_\La(Z)\Big),
~~~~W_\La(Z)=(Z-\oh \SL)\sqrt{Z+\SL},
\eeq
where $V'_{ns}$ denotes the non-scaling part, non-universal part of
$w(z,g)$. $W_\La(Z)$ is usually denoted the continuum 
disk amplitude. Strictly speaking the combinatorial problem
solved assumed that one link was marked. This is equivalent to 
differentiating after $\mi Z$. We denote the disk 
amplitude without marked boundary $\tW_\La (Z)$.
In the same way, differentiating $W_\La(Z)$ after $\mi \La$
is equivalent to having a marked triangle or vertex, i.e. a 
puncture, in the bulk of the disk. We thus obtain:
\beq\label{1.5}
W'_\La(Z)\equiv -\frac{d W}{d\La} \propto \frac{1}{\sqrt{Z + \sqrt{\La}}},
~~~~~   
\tW'_\La(Z)\equiv -\frac{d \tW}{d\La} \propto -\sqrt{Z + \sqrt{\La}}
\eeq
$\tW'_\La(Z,\La)$ is the disk amplitude determined by FZZ  corresponding 
to generalized Neumann boundary conditions and it is related to 
the disk amplitude $W'_\La(Z)$ by differentiation after $\mi Z$,
since the only difference is that the boundary in case of 
the amplitude $W'_\La(Z)$ has a marked point.

The various signs and even zeros of $W$, $W'_\La$ and $\tW'_\La$ might
at first be confusing but are all consistent with the form
\rf{1.4} of the regularized disk amplitude. $w(z,g)$ from 
\rf{1.1} is of course by construction positive and 
the non-scaling part $V'_{ns}$ ensures that this is still the case in 
eq.\ \rf{1.4}. Differentiating $w(z,g)$ a sufficient number of times after 
either $-\La$  or $-Z$  will ensure  that the non-analytic 
part (the scaling part) becomes dominant when the cut-off $a \to 0$.
By construction this dominant part then has to be positive in the 
case of 2d gravity. 

A related remark is that the 
non-positiveness of $W_\La(Z)$ and $\tW'_\La(Z)$
is a reflection of the singular (non-integrable) behavior at zero 
of corresponding partition functions as functions 
of the boundary length $L$:
\beq\label{1.6a}
W_\La(L) = \frac{(1+\SL L)e^{-\SL L}}{L^{5/2}},~~~~  
\tW'_\La(L) = \frac{e^{-\SL L}}{L^{3/2}} , 
\eeq    
where $W_\La(Z)$ and $\tW'_\La(Z)$ are the Laplace transforms 
of $W_\La(L)$ and $\tW'_\La(L)$. In contrast the inverse Laplace transform
of the ``well-behaved'' partition function $W'_\La(Z)$ is integrable
at $L=0$:
\beq\label{1.6b}
W'_\La(L) = \int_{-i\infty}^{i\infty}dZ \; e^{LZ} (Z+\SL)^{-1/2}\propto
L^{-1/2}e^{-\SL L}.
\eeq
 
The various disk partition functions \rf{1.6a} and 
\rf{1.6b} for fixed boundary length $L$ all have the 
common term  $e^{-\SL L}$. 
We can view this term  as an {\it induced} boundary
cosmological term, coming from the bulk cosmological term. 
This is the reason we can actually take $Z$ negative, as long 
as it is larger than $-\SL$.

Let us look at the disk amplitude $W'_\La(Z)$ given by \rf{1.5}. 
It serves without any reservation as the partition function for the 
disk with one puncture. It is positive and one can calculate expectation 
values of the boundary length and the area of the disk 
\beq\label{1.8}
\la L \ra \propto \frac{1}{Z+\SL},~~~ \la A\ra \propto \frac{1}{\SL} \,
\frac{1}{Z+\SL}.
\eeq

There is clearly no trace of non-compactness of the quantum disk 
for $Z= \SL/2$ and the only point where $\la L\ra $ and $\la A\ra$ diverge
is for $Z = -\SL$, minus the induced boundary cosmological constant. 
However, as we shall see below it does
not correspond to (quantum) AdS space in any straight forward way.

\subsection*{Disks with a  geodesic radius}

Obviously, in order to move from a compact space to a non-compact 
space we need a length scale to diverge. 

In quantum gravity 
it is non-trivial to define the concept of length. In 
the seminal work \cite{kkmw} it was shown that proper-time can be defined
constructively in 2d quantum gravity, starting from dynamical 
triangulations. Further, the proper-time propagator was 
constructed, i.e. the amplitude for a universe (a connected loop
of length $L_2$) to be separated a geodesic distance $R$ from 
another universe (a connected loop of length $L_1$). 
In \cite{aw} it was shown that this concept of proper-time or geodesic 
distance allows us to introduce the concept of a divergent 
correlation length for two-point functions in quantum 
gravity and that all the 
standard scaling relations from the theory of critical 
phenomena are still satisfied.
This concept of a divergent correlation length had 
indeed been  missing in 2d quantum gravity until then. 
In \cite{ajw,catterall} and in particular in \cite{aa}
the spin-spin correlation function  
in the Ising model and the 3-state Potts model coupled to 
gravity was shown to  behaved as expected when defined via the 
geodesic distance. In \cite{ope} the first steps were taken 
to define the concept of operator product expansion in terms of 
geodesic distances. 

Let us consider the following cylinder-amplitude: we have a
(marked)  ``entrance loop'' with boundary cosmological constant $Z$ and 
an ``exit loop'' with boundary cosmological constant $Z'$, such that any 
point on the exit loop is separated from the entrance loop by proper-time
(or geodesic distance) $R$, i.e.\ in the path integral we integrate 
over all geometries which satisfy this constraint. As shown in \cite{kkmw} 
the amplitude $G_\La(Z,Z';R)$ satisfies the following equation: 
\beq\label{2.1}
\frac{\prt}{\prt R} G_\La(Z,Z';R) = -\frac{\prt}{\prt Z}
\Big[W_\La(Z) G_\La(Z,Z';R)\Big]
\eeq
with the boundary condition
\beq\label{2.2}
G_\La(L,L';0)= \delta(L-L'),~~~{\rm i.e.}~~~G_\La(Z,Z';0)= \frac{1}{Z+Z'}.
\eeq
The solution is 
\beq\label{2.3}
G_\La(Z,Z';R)= \frac{W_\La(\hZ)}{W_\La(Z)}\; \frac{1}{\hZ+Z'}
\eeq
where $\hZ(R,Z)$ is the solution to the characteristic equation:
\beq\label{2.4}
\frac{\prt \hZ}{\prt R} = -W_\La(\hZ),~~\hZ(R\equ 0)=Z,~~~~{\rm i.e.}~~~~
R= \int^Z_{\hZ} \frac{dX}{W_\La(X,\La)}.
\eeq
Let us shrink the entrance loop to a point (a puncture) \footnote{From 
the general formula for multi-loop correlators $W_\La(Z_1,\ldots,Z_n)$ derived 
in \cite{ajm} it follows that 
$Z^{3/2}_1W_\La(Z_1,\ldots,Z_n) \to -d/d\La( W_\La(Z_2,\ldots,Z_n))$
for $Z_1 \to \infty$.}:
\beq\label{2.5}
W_\La(Z';R)= \lim_{Z\to \infty} Z^{3/2}G_\La(Z,Z';R)= 
\frac{W_\La(\hZ(R))}{\hZ(R)+Z'},
\eeq
where $\hZ(R,Z\equ \infty)$ and $W_\La(\hZ(R))$ are ($ \a= \sqrt{3/2}$):
\beq\label{2.5a}
\hZ(R)=\frac{\SL}{2}\, \Big(1 \pl \frac{3}{\sinh^2(\oh\a\sqrt[4]{\La}R)}\Big),
\eeq
\beq\label{2.5b}
W_\La(\hZ(R))= (\a \sqrt[4]{\La})^3 
\frac{\cosh \oh\a \sqrt[4]{\La} R}{\sinh^3 \oh\a \sqrt[4]{\La} R}.
\eeq
Note that $W_\La(\hZ(R))$ is nothing but the two-point function 
introduced in \cite{aw}.

The function $W_\La(Z;R)$ is similar to the FZZ disk amplitude $\tW'_\La(Z)$
{\it except} that now all points on the boundary have a geodesic distance 
$R$ to the puncture. It is also seen that effectively we have an induced
boundary cosmological constant $\hZ(R)$ in the sense that 
\beq\label{2.6}
W_\La (Z;R) = \int_0^\infty dL \; e^{-ZL}W_\La(L;R),~~~~  
W_\La(L;R)\propto e^{-\hZ(R) L}.
\eeq
When $R$ changes from 0 to infinity the induced boundary cosmological 
constant $\hZ(R)$ changes from the value of the 
boundary cosmological constant at the entrance loop (which we here 
have taken to infinity) to the value corresponding to the zero
of $W_\La(Z)$, i.e. $Z= \SL/2$, as is clear from \rf{2.4}. 

{\it Viewing \rf{2.1} as a renormalization group equation where $R$ is the 
scale factor, we see that it induces a flow in the induced cosmological 
boundary constant and the fixed point corresponding to $R=\infty$ is 
precisely the zero of $W_\La(Z)$.} 

Let us now check how the geometry of the (quantum) pseudo-sphere 
appears from $W_\La(Z;R)$, which serves without any reservation
as an ordinary positive partition function as long as 
$Z$ is larger than minus the induced boundary cosmological
constant $\hZ(R)$. For $R \to \infty$
one obtains (up to corrections $e^{-\a \sqrt[4]{\La}R}$)
\beq\label{2.7}
\la L \ra \propto \frac{1}{Z+\hZ(R)},~~~~~
\la A\ra \propto \frac{1}{\SL(Z+\hZ(R))}.
\eeq
This expression looks very similar to \rf{1.8} derived for  the 
disk amplitude $W'_\La(Z)$, the bulk-induced boundary cosmological 
constant $\SL$ being replaced with $\hZ(R)$. However, precisely for  
$Z=-\hZ(\infty)=-\SL/2$ we obtain, using \rf{2.5a}
\beq\label{2.8}
\la L \ra \propto \La^{-1/2}e^{\a\sqrt[4]{\La}R} ,~~~~~
\la A\ra \propto \La^{-1}e^{\a\sqrt[4]{\La}R}
\eeq
For a fixed $\La$ we thus have indeed the features of 
the pseudo-sphere when $R \to \infty$: The circumference and the 
area are proportional and go to infinity exponentially in the 
``radial'' geodesic distance. The only difference is 
that the dimension of the geodesic distance $R$ is anomalous (for reasons
which will be explained in the next section). This anomalous scaling 
of the geodesic distance is a ``quantum
geometry'' aspect of 2d quantum gravity with 
$c_{matter} \equ 0$ (or $c_{Liouville} \equ 26$ in the corresponding 
Liouville theory) and is a 
manifestation of the fractal structure of a typical 
geometry: it has Hausdorff dimension 4, as emphasized in 
\cite{kkmw} and \cite{aw}.

\subsection*{The ordinary disk amplitude}

\begin{figure}
\centerline{\hbox{\psfig{figure=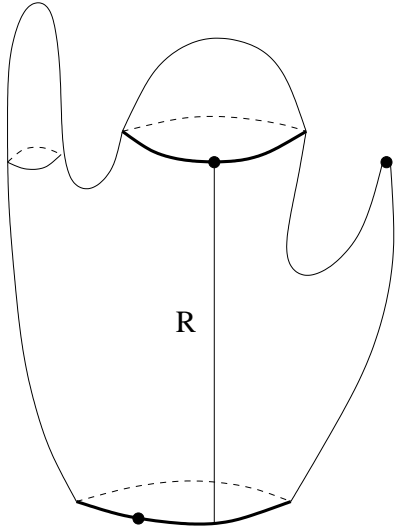,height=7cm,angle=0}}}
\caption[identity]{Marking a vertex in the bulk of $W_\La(Z)$. The mark
has a distance $R$ from the boundary loop, which itself has one marked vertex.
Shown on the figure are also two other loops with the 
same geodesic distance to the entrance loop, the loop to the right
being of microscopic size at the very tip of the right ``branch''.}
\label{identity}
\end{figure}

Let us return to the ordinary disk amplitude given by the 
partition function $W'_\La(Z)$. If we want to relate 
to the work of ZZ we have to keep $\La$ fixed. The only 
way to get a divergent boundary length and a divergent area
is to tune $Z$ to $\mi \SL$, minus the induced boundary cosmological 
constant. Let us calculated the average geodesic distance 
to a point in the interior of the disk. This calculation
is a nice example of the use of ``quantum geometry'' in the 
context of the path integral and 
was first used in a slightly different context in \cite{al}.
From Fig.\ \ref{identity} it is clear that we have the following identity 
before the cut-off is removed
\beq\label{an2}
g\ \frac{\prt w(x,g)}{\prt g}=
\sum_r \oint \frac{dz}{2\pi i} \;G(x,z^{-1};g;r)\; 
\frac{\prt (z w(z,g))}{ z\,\prt z}.
\eeq 
The lhs is the disk amplitude with one marked point. A given mark has a 
distance $r$ ($R$ in the continuum) to the entrance loop and it lies 
on a connected loop which has the distance $r$ to the entrance loop  
\footnote{Note that there can be many 
disconnected loops at the same distance.}. If the loop 
passing through the marked point has length $l$ the loop is  closed 
by a cap: the disk amplitude $w(l,g)$ itself, which can 
be glued in $l$ ways to the marked loop of the same length.
This decomposition of the disk amplitude, summing over $r$ and $l$,
is implemented on the rhs of eq.\ \ref{an2}.  
As shown in \cite{al} \rf{an2} has 
precisely two solutions \footnote{By solution we mean 
a simultaneous determination of $G(x,z;r)$ and the scaling 
part of the function w(z,g).} which also have a scaling limit: 
one which does {\it not} allow a spatial 
loop to split into two under the proper-time evolution governed 
by time $R$ and where the underlying continuum 
theory  in \cite{al} was called {\it Lorentzian quantum 
gravity} (since each geometry had an interpretation as 
a causal geometry after rotation to Lorentzian signature), 
and one where the underlying 
continuum theory is the 2d {\it Euclidean quantum gravity} discussed here
(Liouville quantum gravity).  

Here we will be interesting the 2d Euclidean quantum gravity 
solution to \rf{an2}. Thus inserting \rf{1.4} in \rf{an2}
and extracting the leading singular behavior on the lhs of 
\rf{an2}
we obtain first that the geodesic distance {\it has} to scale 
anomalously ($R= r \sqrt{a}$), as already used, and next  
\beq\label{3.1}
W'_\La(Z) \propto \int_0^R dR \int_{-i\infty+c}^{i \infty +c} dZ' \; 
G_\La(Z,\mi Z';R)=
 \int_0^\infty dR \;G_\La(Z,L\equ 0), 
\eeq
where $c > Z$ and where, as follows from \rf{2.3},
\beq\label{3.1a}
G_\La(Z,L\equ 0;R)= \frac{W_\La(\hZ(R,Z))}{W_\La(Z)}.
\eeq

At first sight eq.\ \rf{3.1} might be  a little surprising since any reference 
to $W_\La(Z)$ has disappeared in the rhs of \rf{3.1}, the non-scaling 
part of $w(z,g)$ giving the leading contribution. The explanation is 
the singular behavior of $L W_\La(L) = L^{-3/2}+\cdots$ for small $L$.
Thus, picking an arbitrary point in the bulk, the chance that the 
loop is of cut-off length $a$ is totally dominant, i.e.\ the non-scaling 
part of $w(z,g)$ comes into play. In this sense Fig.\ \ref{identity}
is misleading: the loop in the bulk should be infinitesimal, like the 
loop indicated on the right ``branch'' in Fig.\ \ref{identity}.

Eq.\ \rf{3.1} shows that the ``radial'' distribution for the 
disk amplitude $W'_\La(Z)$ is $F_Z (R) = G_\La(Z,L\equ 0;R)$, and
we obtain
\beq\label{3.2}
\la R \ra_Z =\frac{1}{W'_\La(Z)} 
\int_0^\infty dR \;F_Z(R)\, R \sim \frac{1}{\sqrt[4]{\La}},
\eeq
even if $Z \to -\SL$. 
How is it possible to have both $\la A\ra \sim \la L \ra \to \infty$
and the geodesic distance between any points finite at the same 
time: only if the disk looks nothing like a ``nice'' disk in that limit 
but is highly fractal.

\subsection*{Discussion}

We have shown how the special points $Z^2 = \La/4$ are related to  
a divergent length scale in 2d Euclidean quantum gravity
and we have shown how one can obtain non-compact ``quantum'' pseudo-spheres 
associated with these values of the boundary cosmological constant 
\footnote{Notice that precisely the relation $Z^2 = \La/4$ actually 
appears in the work of FZZ (eq.\ (3.5), as the special points
where certain boundary operators satisfy a particular simple 
equation. This is of course much better understood now, by i.e.\
the work \cite{ss}}.

\vspace{12pt}
\noindent
Let us end with some comments: 
\vspace{12pt}

\noindent
(1) The combinatorial solution \rf{1.4} of Tutte is easily generalized  
to the so-called multi-critical models which correspond to 
$(p,q)\equ (2,2m \mi 1)$ 
minimal conformal models coupled to 2d gravity, $m > 2$
($m\equ 2$ is the $c\equ 0$ situation considered above):
\beq\label{1.7}
w(z,g)=\oh \Big(V'_{ns}(z)+a^{m-1/2}W^{(m)}_\La(Z)\Big),~~~
W_\La^{(m)}(Z)=M_{m-1}(Z)\sqrt{Z+\SL}.
\eeq
$V'_{ns}(z)$ and $M_{m-1}(Z)$ are  polynomials of  order $m$ and 
 $m \mi 1$, respectively, $V'_{ns}(z)$ being the 
non-scaling part of $w(z,g)$. In \cite{ss} is was shown that for a given  
value of $\La$ the ZZ-branes 
were associated with the values of $Z$ where $M_{m-1}(Z)=0$ 
\footnote{And they generalized
this to arbitrary (p,q) minimal conformal models coupled to 2d gravity}.
In \cite{gk} it was shown that one has an equation similar to \rf{2.1}
only with $W_\La(Z)$ replaced by $W^{(m)}_\La(Z)$. Clearly,
one has a divergent length scale associated with each of the zeros 
of $M_{m-1}(Z)$. However, the interpretation of $R$ as a geodesic 
distance is no longer straight forward. Rather, different divergent 
$R$'s are associated with the different critical behavior of 
the matter coupled to gravity for $c \neq 0$, but are 
of course not unrelated 
to the geodesic distance. As mentioned we will discuss this elsewhere
\cite{aagk2}.

\vspace{12pt}
\noindent
(2) The observation by SS that the zero 
of $W_\La(Z)$ is related to ZZ-branes is at first very surprising:
If we expand in powers of $1/Z$ we obtain
\beq\label{4.1}
W_\La(Z) = Z^{3/2}-\frac{\La}{Z^{1/2}} + c \frac{\La^{3/2}}{Z^{3/2}} +\cdots 
\eeq
By inverse Laplace transformation with respect to $\La$ it is seen that the 
first two terms correspond to the {\it zero} area limit: one obtains
$\del (A)$ and $\del'(A)$ \footnote{In fact, differentiating \rf{4.1} after 
$\mi \La$ the term $1/Z^{1/2}$ has the interpretation as partition
function for the so-called branched polymers with two marked points:
no interior, only the boundary which can be viewed as a 
planar branched polymer
\cite{adf,adfo}.}. Thus, when discussing the relationship between 
matrix models and integrable systems, often one only considered 
the negative powers of $W^{(m)} (Z)$ starting at $Z^{-3/2}$. 
If we consider this part of \rf{4.1} it is explicitly positive for 
all positive values of $Z$. 

We now understand that because of the flow of the induced boundary 
cosmological constant as a function of the scale $R$ of the 
universe, indeed the zero of $W_\La(Z)$ plays an important 
physics role, namely as the fixed point for the induced 
boundary cosmological constant for $R \to \infty$.

\vspace{12pt}
\noindent
(3) The geodesic distance scales anomalously in 2d quantum gravity.
This is related to the ``emission'' of a large number of 
microscopic baby universes if we view $R$ as proper-time and 
the two-loop function as the propagator of spatial loops in 
proper-time \cite{kkmw}. It is possible to ``integrate out'' the 
baby universes \cite{ackl} and in this way one is left precisely 
with propagator of 2d Lorentzian quantum \cite{al}. For this 
propagator one can repeat the arguments  presented above,
and one finds now  instead of \rf{2.8}
the real formulas for pseudo-sphere: the geodesic distance
now scales canonically \footnote{Details will be presented elsewhere
\cite{alnew}.}, so in a quite precise way the quantum pseudo-sphere
discussed here is the (extremely) hairy version of the ordinary 
pseudo-sphere.

\section*{Acknowledgment}

J.A. was supported by ``MaPhySto'', 
the Center of Mathematical Physics 
and Stochastics, financed by the 
National Danish Research Foundation.
S.A. would like to thank the Niels Bohr Institute for kind hospitality.

\end{document}